# Reconstruction of the lattice Hamiltonian models from the observations of microscopic degrees of freedom in the presence of competing interactions

Mani Valleti,[1,a] L. Vlcek,[2,3] Maxim Ziatdinov,[4] Rama K. Vasudevan,[4] and Sergei V. Kalinin[4,b]

[1] Bredesen Center for Interdisciplinary Research, University of Tennessee, Knoxville, TN 37996, USA

[2] Joint Institute for Computational Sciences, University of Tennessee, Knoxville, Oak Ridge, TN 37831, USA

[3] Materials Science and Technology Division, Oak Ridge National Laboratory, Oak Ridge TN 37831, USA

[4] The Center for Nanophase Materials Sciences, Oak Ridge National Laboratory, Oak Ridge, TN 37831

The emergence of scanning probe and electron beam imaging techniques have allowed quantitative studies of atomic structure and minute details of electronic and vibrational structure on the level of individual atomic units. These microscopic descriptors in turn can be associated with the local symmetry breaking phenomena, representing stochastic manifestation of underpinning generative physical model. Here, we explore the reconstruction of exchange integrals in the Hamiltonian for the lattice model with two competing interactions from the observations of the microscopic degrees of freedom and establish the uncertainties and reliability of such analysis in a broad parameter-temperature space. As an ancillary task, we develop a machine learning approach based on histogram clustering to predict phase diagrams efficiently using a reduced descriptor space. We further demonstrate that reconstruction is possible well above the phase transition and in the regions of the parameter space when the macroscopic ground state of the system is poorly defined due to frustrated interactions. This suggests that this approach can be applied to the traditionally complex problems of condensed matter physics such as ferroelectric relaxors and morphotropic phase boundary systems, spin and cluster glasses, quantum systems once the local descriptors linked to the relevant physical behaviors are known.

---
[a] Corresponding author, svalleti@vols.utk.edu
[b] Corresponding author, sergei2@ornl.gov

Phase transition between dissimilar phases underpin multiple areas of condensed matter physics and materials science. From applications viewpoint, phase transitions are intricately involved in the process of materials formations in virtually all technologies from ceramics to metals to semiconductors. Electrical, gating, and chemical control of phase transitions now emerges as a promising paradigm for the low-voltage electronics in tunneling barriers and field effect devices.[1, 2] Beyond condensed matter physics, phase transitions in Bose-Einstein condensates allow new classes of quantum information technology devices.[3]

The ubiquity of the phase transitions in virtually all areas of applied and fundamental science have made them one of the central areas for the theoretical exploration. Among the existing paradigms for exploring phase transitions, a special place is held by the lattice models such as Ising[4], Heisenberg[4], Kitaev[5], etc. Here, the system is represented as a set of interacting spins on a spatial lattice. The spins represent the individual degrees of freedom of material and can be magnetic spins, electrical dipoles, atomic species, etc. The lattice defines the geometry of the interactions. Depending on the conservation laws, character of the spins, interactions, and geometry, these lattice models can represent an extremely broad range of phenomena from ordering in metal alloys to ferroic transitions to surface adsorbate dynamics.

The universality of the lattice model approach has spurred intensive analytic and numerical investigations of these models.[6, 7] These studies mainly seek to explore the phase diagrams of the lattice models, i.e. correspondence between the global variables such as temperature and field and preponderant behaviors and patterns for spin arrangement. Similarly of interest are global thermodynamics parameters such as magnetization, heat capacity, and susceptibility. Finally, the spatial arrangements of the spins are traditionally explored in the form of the ground state pattern, as well as correlation functions representing the distributions of the defects and disorder, i.e. over which length scales the system is ordered. Notably, these behaviors allow for straightforward exploration only for systems with well-defined ground states, whereas the presence of the competing interactions leads to frustrated and poorly defined ground states. Nonetheless, the combination of efficient sampling strategies and high-performance computing have allowed exploration even complex multidimensional models.

Over recent years, the emergence of machine learning (ML) tools and readily availability of information-compression algorithms, such as principal and independent component analysis, support vector machines, as well as deep learning tools such as variational autoencoders have provided a new impetus of the field.[8-11] Here, we apply ML techniques to suitably chosen local variables to compress the spatial spin arrangements to a small number of control variables that can be identified with the function of controlling order parameters. Even though the applications of ML tools to lattice models are relatively recent, the field has experienced an exponential growth in the last 3 years, opening pathways to exploration of exotic quantum phases, off-lattice models, etc.[12-15]

However, experimental selection of the lattice model corresponding to a specific phase transition and determination of its parameters represent a largely unresolved problem. Traditionally, the symmetry and type of order parameter are determined based on the macroscopic measurements and scattering studies, whereas numerical values of exchange integrals are obtained based on the matching between theoretical predictions and macroscopic measurements of thermodynamic parameters. However, these studies typically require high quality samples for which the corresponding critical behaviors can be reliably determined. Similarly, in the materials with high disorder, non-ergodic or non-thermalized ground states determination of appropriate lattice models can be highly non-trivial.

The development of high-resolution imaging tools such as scanning transmission electron microscopy and scanning probe microscopy have allowed direct insight into the atomic-scale structure of materials. Beyond atomically resolved and element specific imaging, these techniques provide the information on the minute symmetry breaking distortions, allowing for direct mapping of the order parameter fields such as polarization, octahedra tilts, and strains. Correspondingly, of interest is the extraction of the parameters of the underpinning physics model from the atomically resolved observations. As one approach, this can be achieved using the direct matching of the mesoscopic order parameter field to the Ginzburg-Landau type models. Alternatively, the mesoscopic descriptors such as hysteresis loops or relaxation curves can be used as a proxy identifiers for local behavior. Finally, it has been proposed that direct observation of the mesoscopic degrees of freedom can be directly compared to lattice model via statistical distance minimization[16, 17].

Previously, we have reported the principles of the statistical distance minimization, as well as applications for establishing of the parameters of the non-ideal solid solutions in layered superconductors, information fusion between surface and depth-resolved chemical information in manganites, and pair- and triple-atom model comparison for segregation in layered chalcogenides[18]. We have also explored the veracity of the statistical distance minimization and associated uncertainty quantification for the paradigmatic Ising model and demonstrated that the in the presence of weak bond-disorder the exchange integral can be determined well above associated bulk phase transitions.

Here, we explore the reconstruction of the Ising model parameters in the presence of the competing interactions that can give rise to the frustrated ground states and explore the sensitivity of the reconstruction to the incomplete knowledge of the model. The guidelines for experimentally driven model selection are proposed.

**1. Model and phase diagram**

Here we explore the behavior of the Ising model with the nearest neighbor and next nearest neighbor interactions. Depending on the sign and magnitude of the exchange integrals, this model can give rise to a rich phase diagram containing ferromagnetic, antiferromagnetic, paramagnetic, and frustrated phases. Here, first we apply the ML tools for rapid simulation of the associate phase diagram based on the local statistics.

**1.1. Model and simulation**

The classical Ising Hamiltonian model realized on $N^2$ -evenly spaced lattice sites of a square lattice with nearest and second nearest neighbor interactions is the chosen model for this study. The Hamiltonian of a given configuration $\sigma$ in the absence of external magnetic field is given by equation (1).

$$H(\sigma) = - \sum_{<i,j>} (J_1 + \Delta J_1)\sigma_i\sigma_j - \sum_{<i,k>} (J_2 + \Delta J_2)\sigma_i\sigma_k \qquad (1)$$

where $J_1$ and $J_2$ are the exchange integrals corresponding to the nearest neighbor interactions and next nearest neighbor interactions respectively, $\Delta J_1$ and delta $\Delta J_2$ are the corresponding disorders in exchange integrals, <i,j> and <i,k> are the sums over all nearest neighbors and next nearest neighbors respectively.

A Monte Carlo (MC) simulation was performed on a square lattice with $N = 40$ using the Metropolis algorithm. The system was equilibrated for $1000 \times N^2$ Monte Carlo steps and the data is acquired on the next $1000 \times N^2$ Monte Carlo steps. A spin configuration reversal of each lattice site was attempted at every step. The flip was accepted when the energy of the resultant configuration was lower than the previous step. Else, the probability of spin flip was estimated using the Boltzmann distribution given by equation (2), where $\beta$ is the inverse temperature and denominator is the partition function.

$$P_\beta(\sigma_i) = \frac{e^{-\beta H(\sigma_i)}}{\sum_j e^{-\beta H(\sigma_j)}} \tag{2}$$

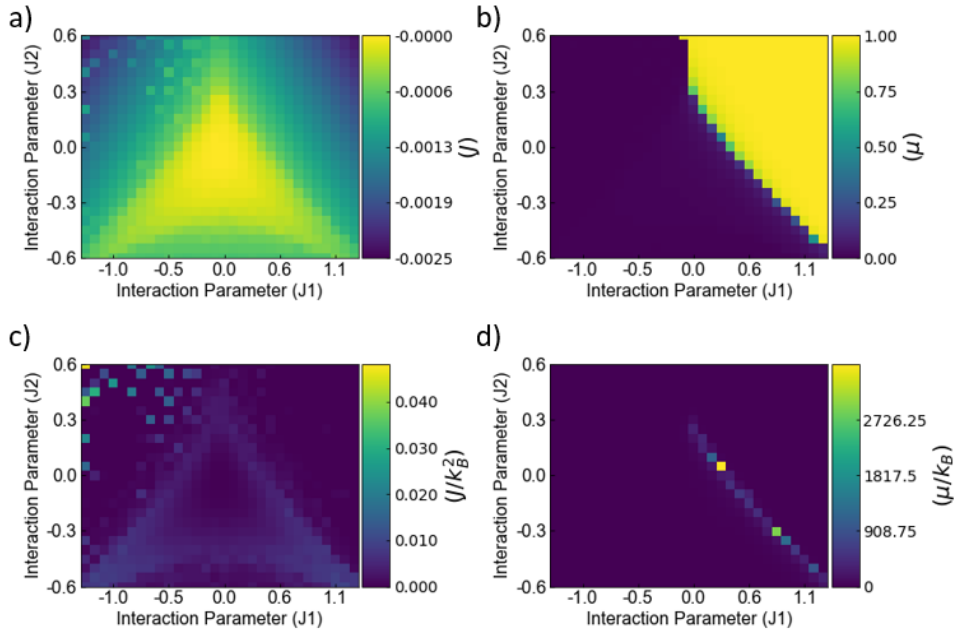

**Figure 1.** a) Energy, b) Magnetization, c) Specific Heat and d) Susceptibility of the square lattice Ising model at reduced temperature Tr = 0.96 as functions of exchange integrals

Shown in Figure 1 are the classical thermodynamic properties plotted as a function of the interaction parameters. Figure 1 (a) illustrates the energy per site. The corresponding magnetization is shown in Figure 1 (b), clearly highlighting the ferromagnetic region and the transition between the ferromagnetic and paramagnetic states (diagonal line). The specific heat is shown in the Figure 1 (c), delineating the primary phase transitions between ferromagnetic, antiferromagnetic, paramagnetic, and frustrated states. Finally, susceptibility is shown in Figure 1 (d), showing the divergence across ferromagnetic – paramagnetic phase transition. Overall, Figure 1 provides a traditional approach for mapping the phase diagram of lattice models based on the macroscopic thermodynamic parameter analysis.

**1.2. Machine learning analysis of the phase diagram**

An alternative approach for mapping phase diagrams in simulated lattice models is based on the machine learning and statistical analysis of the local spin configurations and correlation between spins. For example, multiple works showed that using a supervised learning techniques such as convolutional neural networks, which are traditionally used for image recognition tasks, one can identify a phase transition and map phase boundaries in the classical and quantum lattice models using 2-dimensional and 3-dimensional "images" of the spin configurations around the critical points as a training set[8-10]. At the same time, the more conventional multivariate statistics and probabilistic learning tools such as kernelized principal component analysis and variational autoencoder allowed learning a phase transition and the associated order parameter in the classical Ising model from local spin configurations in the unsupervised manner[19].

Traditional descriptors of such microstates are spin configurations of all the lattice points[20]. In this scenario, the number of rows ($R$) in the feature matrix are $R = N_{sim} * N_{mc}$. While columns ($C$) are the number of features used for the analysis. Spin configurations of the entire lattice at each simulation are used as attributes. The first few principal components corresponding to such matrix closely resemble the macroscopic properties (magnetization, energy, specific heat and susceptibility) of the system at a given set of simulation parameters.

Here, we propose the approach where phase classification is performed based on the relative frequencies of local configurations at a given set of simulation parameters. These frequencies of local configurations describe the microstates of the system exhaustively[17]. For a given case, these are averaged over all the microstates generated by the Monte Carlo simulations. The local configurations considered for further analysis are described in Fig 2e. The relative frequencies of local configurations of ferromagnetic, anti-ferromagnetic, paramagnetic and the frustrated systems are shown in Fig 2a-d. In this case, the number of elements of the feature matrix are 5400 as opposed to $10^9$ in the study referred. Monte Carlo simulations are performed at a reduced temperature of $T_r = 0.96$ and at 900 evenly spaced points on a grid space of exchange integrals $J_1 = [-1.25, 1.25]$ and $J_2 = [-0.75, 0.75]$. Phase diagrams of complex systems can be mapped using statistical analysis of the local structures values.

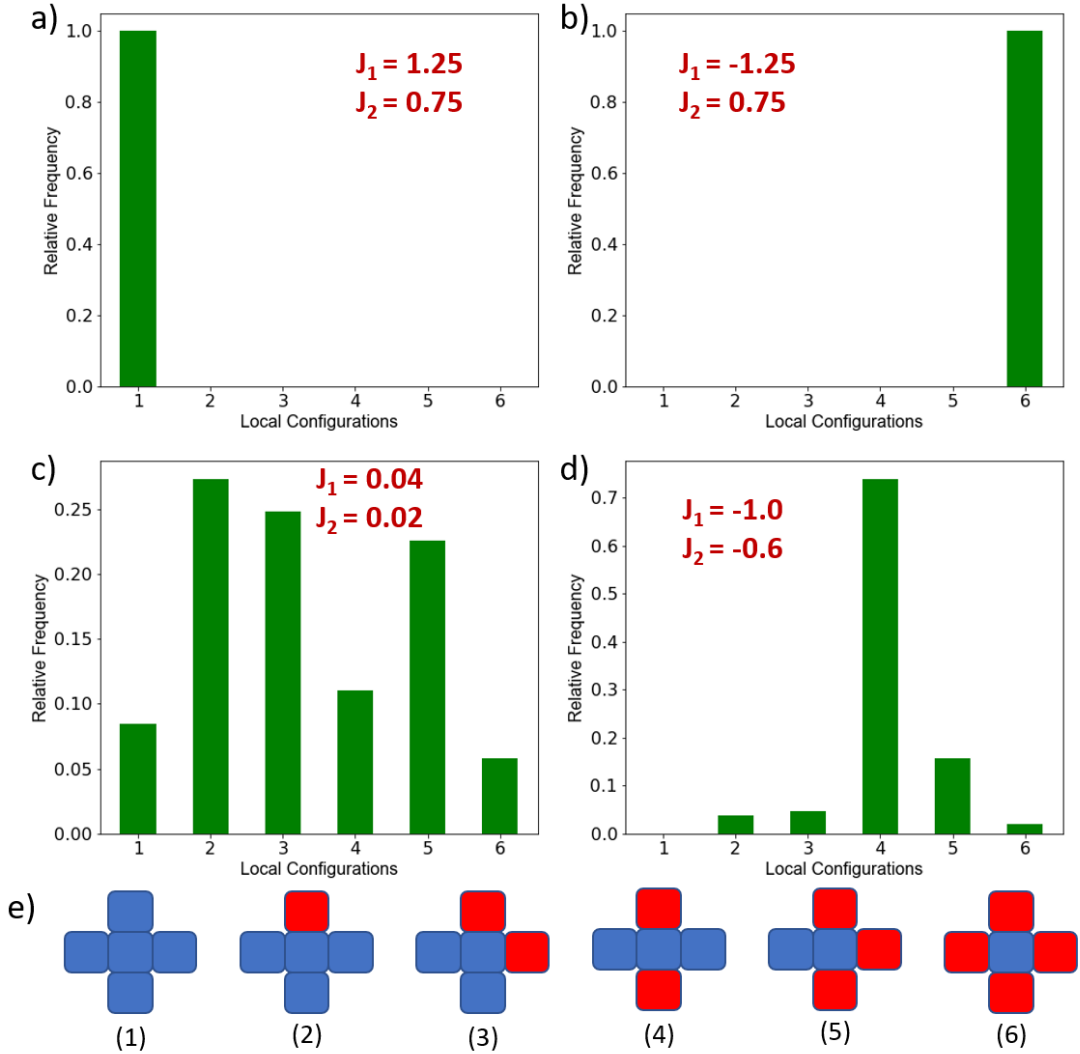

**Figure 2.** Representative relative frequencies of local configurations at reduced temperature $T_r = 0.96$ for (a) paramagnetic, (b) anti-ferromagnetic, (c) paramagnetic and (d) frustrated regimes of phase diagram. (e) Exhaustive set of six local configurations which serve as descriptors for a given microstate. Representative configurations for each regime are show in Fig. 4d in the same order.

To establish the phase diagram of the system over the parameter space, we use simple k-means clustering approach. A similar analysis was proposed by Canabarro et al. in which the pairwise correlations among all spins are used as the descriptors[21]. This algorithm segregates the input data into k- clusters[22]. Mean of each cluster is used as a cluster identifying feature and each observation is assigned to the cluster with the nearest mean. k-Means minimizes variances within clusters and is given by the equation 3.

$$\underset{S}{argmin} \sum_{i=i}^{k} \frac{1}{2|S_i|} \sum_{x,y \, \epsilon \, S_i} ||x-y||^2 \qquad (3)$$

The number of clusters is a priori unknown but can be estimated based on the quality of separation and the target output. Here, when the feature matrix is segregated into 4 clusters using the k-Means clustering technique, the resulting clusters (Fig. 3a) exactly resembled the energy (Fig. 1a) and specific heat (Fig. 1c) phase diagram. It is also observed that for k = 2, the clustering plot reproduced the magnetization and susceptibility phase diagrams. Representative configurations for the four clusters are plotted in Fig. 3d. The clusters are also plotted in two-dimensional space of first two principal components[22] (Fig. 3b) and as a dendrogram (Fig. 3c) to show the hierarchy between the clusters. The subsequent increase of the clusters number leads to the emergence of additional regions concentrated at the boundary between frustrated and paramagnetic states; however, primary ferromagnetic, antiferromagnetic, and frustrated regions remain invariant.

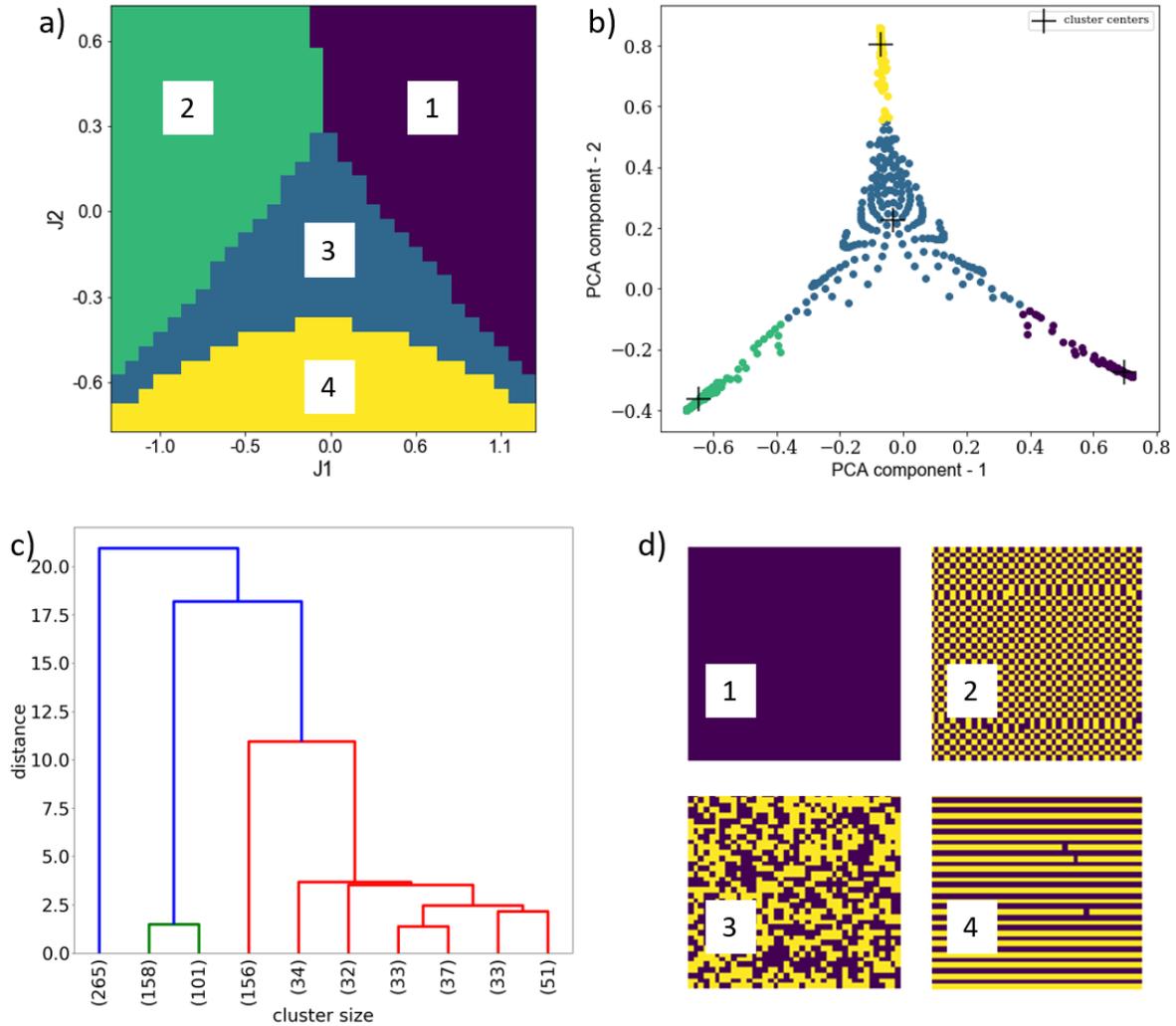

**Figure 3.** (a) Four clusters generated as a result of k-Means algorithm when relative frequencies of local configurations are used as descriptors of macro states. (b) Clusters and their centers plotted in 2-D space of first two principal components. (c) Dendrogram generated using hierarchical clustering. (d) Representative configurations that correspond to the four clusters formed.

We note that classically, the exploration of the lattice models relies on the macroscopic thermodynamic descriptors such as shown in Figure 1, as well as structure factors and correlation functions. Here, we rapidly map the relevant phase diagram using the comparison of the local motifs. In addition to the obvious and previously reported applications for exploration of the phase diagrams of the lattice models, these observations suggest that the statistics of local spin configurations contain the information on the microscopic interactions in Eq. (1).

**2. Reconstruction of interaction from microscopic degrees of freedom**

Here, we extend the statistical distance minimization method to reconstruct the interaction in the Ising model within the full parameter space. During the reconstruction, we have access to the spin configurations (synthetic experimental data), but not the Ising model parameters (value of exchange integrals) per se. Correspondingly, the goal of the reconstruction is to determine the exchange integrals (parameters of the generative physics model) based on the observations of local configurations, and establish the associated uncertainties.

To determine the unknown model parameters, we run the Monte Carlo simulations at a given set of simulation parameters to generate the spin configurations of the base (pseudo-experimental or synthetic experimental data) case. Here, twenty snapshots from the last 400 steps of the MC simulations are used as experimental observables for the base case. Simulations were carried out on the entire grid space of the exchange integrals to reconstruct the exchange integrals of a given base case. The local configurations of the base case are then compared to the simulated cases using the statistical distance metric[23]. This metric is employed to quantify the distinction between pair thermodynamics systems and is given by

$$s = \arccos\left(\sum_{i=1}^{l} \sqrt{p_i}\sqrt{q_i}\right) \qquad (3)$$

where $p_i$ and $q_i$ are the probabilities of finding a local configuration $i$ in the base case and model simulations respectively, and the summation runs over all possible (6 in this case) local configurations. Statistical hypothesis testing indicates that as the measured distance metric reduces to zero, the target and model system measurements become indistinguishable. A unique characteristic of this metric, which sets it apart from popular measures such as Kullback-Leibler divergence, is that it automatically incorporates the statistical weights of the collected data, allowing thus for separation of weak signals (viz. thermal fluctuations) from the statistical noise inherent in samples of a thermodynamic system. Extracting these fluctuations from the data is critical for improving the predictive capabilities of the optimized model, as they encode the system's response to the changes of external conditions.

## 2.1. Reconstruction of the full $J_1$ - $J_2$ model

First, we explore the reconstruction of the model parameters for the full model. The reconstruction results for all four quadrants of the phase diagram are shown in Figures 4 and 5. For all values of $J_1$ and $J_2$, the reconstruction shows that the statistical distance minimization yields unbiased reconstruction for the model parameters. Here, two base cases ($J_1$ = +0.75, $J_2$ = +0.35 and $J_1$ = -0.75, $J_2$ = +0.35) were reconstructed that have a well-defined ground state where the nearest and next nearest neighbor interactions are not frustrated. The uncertainty is quantified using the experimental averaging technique discussed in the earlier studies[17].

The reconstruction fails below the phase transition region as there is no information in the spin configurations. Exchange integrals were reconstructed with high certainty at temperatures just above the phase transitions, where the configurations are strictly defined by the interaction parameters. The uncertainty of reconstruction increases with increasing temperature due to random spin configurations. Still, the reconstruction is possible for temperatures that are an order greater than the transition temperature before the uncertainty impedes the interpretation.

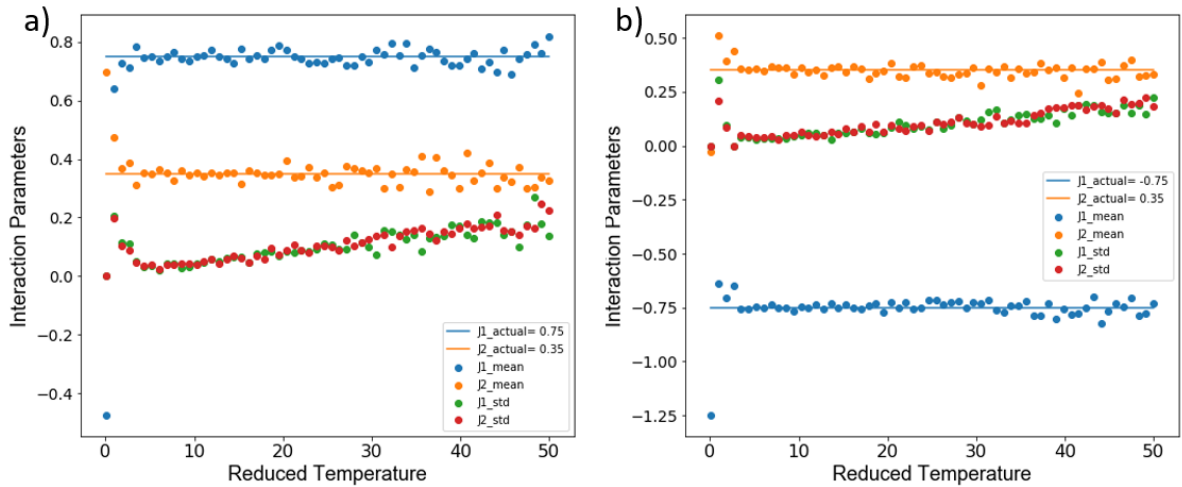

**Figure 4.** Reconstructed values of the exchange integrals and their uncertainties in the (a) first ($J_1$>0, $J_2$>0) and (b) second ($J_1$<0, $J_2$>0) quadrants, where there are well defined ground states, as a function of reduced temperature.

For the next two cases ($J_1 = -0.75$, $J_2 = -0.35$ and $J_1 = +0.75$, $J_2 = -0.35$), the nearest and next nearest neighbor interactions are incompatible and lead to the frustrated ground states. The ground states of these systems are not well defined. Proposed reconstruction technique works even in this scenario (of no well-defined ground states). Since there is no apparent transition in these regions, the reconstruction works even at low temperatures (Figure 5). High temperature behavior for these cases shows similar behavior to that described for the first two cases.

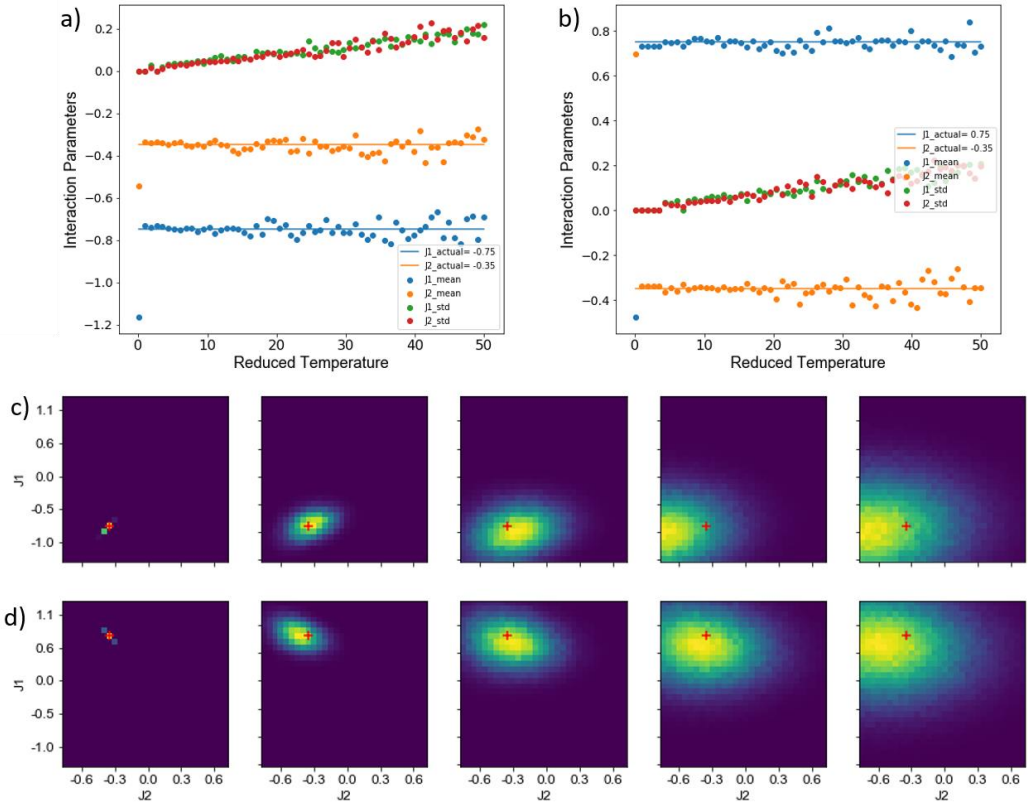

**Figure 5.** Reconstructed values of the exchange integrals and their uncertainties in the (a) third ($J_1<0$, $J_2<0$) and (b) fourth ($J_1>0$, $J_2<0$) quadrants, where there are well defined ground states, as a function of reduced temperature. Likelihoods of the reconstructed values in the exchange integrals space of the (c) third and (d) fourth quadrants. The values of the exchange integrals used to produce to the pseudo experimental data is shown by the cross in each plot (c and d).

To complement the analysis of the uncertainties via the corresponding point estimates of mean and dispersion and the marginalized posterior distribution for individual exchange integral values, we also analyze corresponding joint distributions to perform uncertainty quantification (UQ) in the 2D parameter space. The log-likelihood of a sample at a distance $s$ from the limiting probability distribution is then proportional to $-2ns^2$, where $n$ is the number of samples (individual local configurations)[17, 24]. As the distance $s$ increases the log-likelihood decreases and it does so with a linear dependence on the number of samples $n$. The likelihood over the exchange integral space is shown in Fig. 5c-d at different points of reduced temperature. The increase in uncertainty with temperature can be visualized as the increase in spread of likelihood.

**2.2. Reconstruction for overdetermined and underdetermined model**

In this section, we explore the cases where the pseudo experimental (base) case is guided only by the interactions between the nearest neighbors ($J_1$ =0.75) and is reconstructed by simulations incorporating both nearest and next nearest neighbors ($J_1$ and $J_2$). This case is equivalent to the case where $J_2$ (the exchange integral corresponding to the second nearest neighbors) is zero. The reconstructions follow a similar trend as discussed in the previous cases (Figure 5a). It works with very high certainty post the transition temperature and the uncertainty increases with temperature. Interestingly, for this case the zero value of the $J_2$ can be reliably determined.

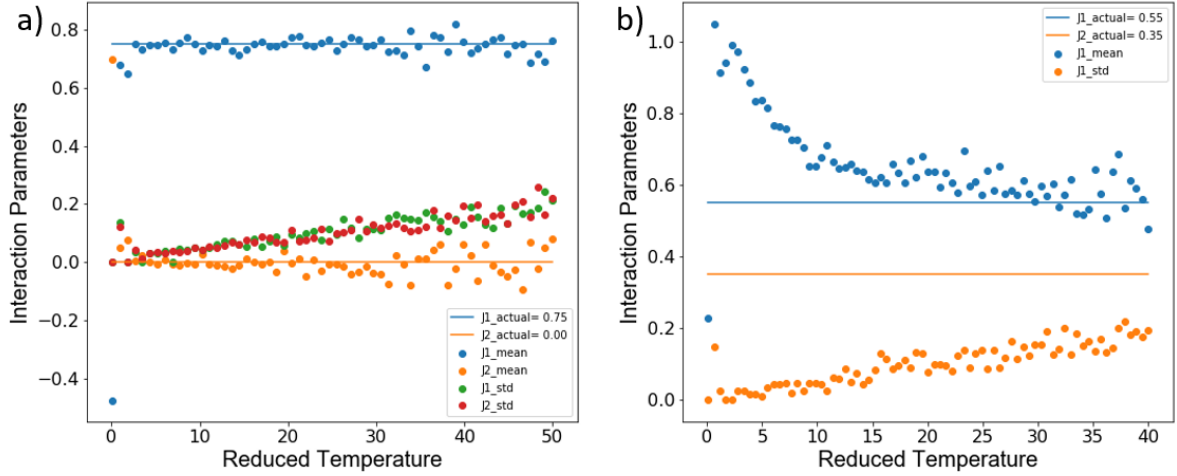

**Figure 6.** Reconstructed values of the exchange integrals and their uncertainties in the (a) overdetermined and (b) underdetermined cases as a function of reduced temperature.

We further explore the reconstruction of an underdetermined model (Figure 6b) where the experimental image is guided by both nearest and next nearest neighbors ($J_1$= 0.75, $J_2$ = 0.35) while it is tried to be modeled using the simulations with nearest neighbor interactions ($J_1$). The model tries to fit the data by overestimating the exchange integral, but the estimated value is different at different temperatures. It can be inferred from the reconstruction curves that a model with only the nearest neighbor interactions will not be able to reproduce the experimental results that are governed by both nearest and next neighbors. The likelihood for both these cases in the exchange integral space is shown in Fig7 a and b respectively.

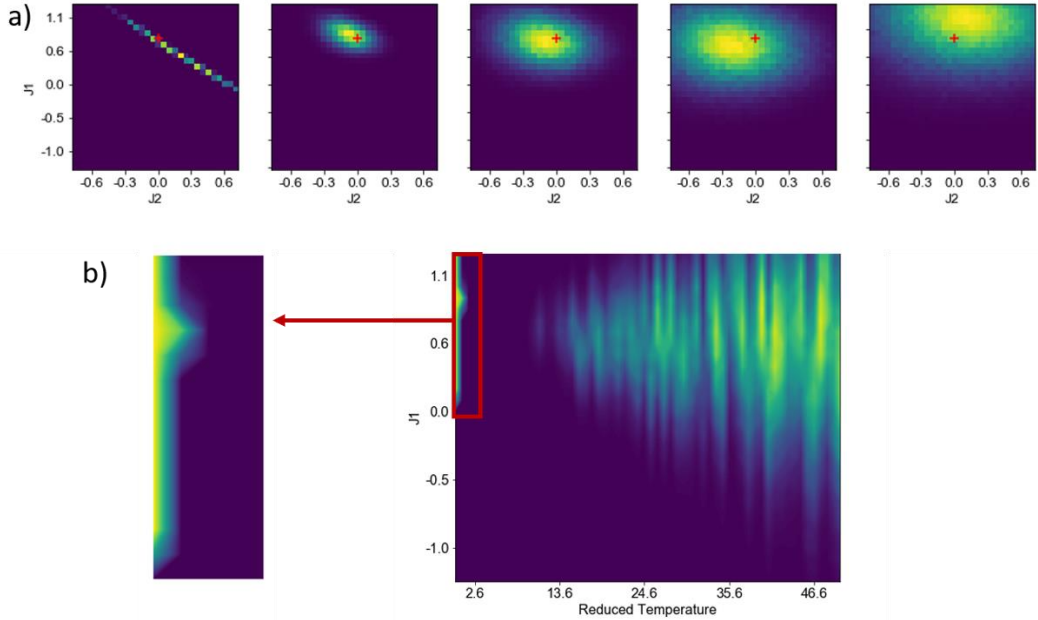

**Figure 7.** Likelihoods of the reconstructed values in the exchange integrals space of (a) overdetermined and (d) underdetermined cases. The values of the exchange integrals used to produce to the pseudo experimental data in the overdetermined case is shown by cross in (a) and the value of exchange integrals used in underdetermined case are $J_1 = 0.55$ and $J_1 = 0.35$.

## 3. Bond-disorder effects on the uncertainty quantification

Reconstruction until this section has been performed on a square lattice in which the values of exchange integrals in the pseudo experimental data set were constant over the entire lattice. In this section, we explore the reconstruction when a bond-disorder is introduced in the exchange integrals. To execute this, every lattice site is associated with a different value of exchange integral $J+\Delta J$, where $\Delta J$ is normally distributed around $J$ with a specified standard deviation. The disorder is added to both exchange integrals $J_1$ and $J_2$. The standard deviation is referred to as disorder for the rest of the section. The bond disorder was only introduced in the pseudo-experiment case while the simulation cases were run at $\Delta J = 0$.

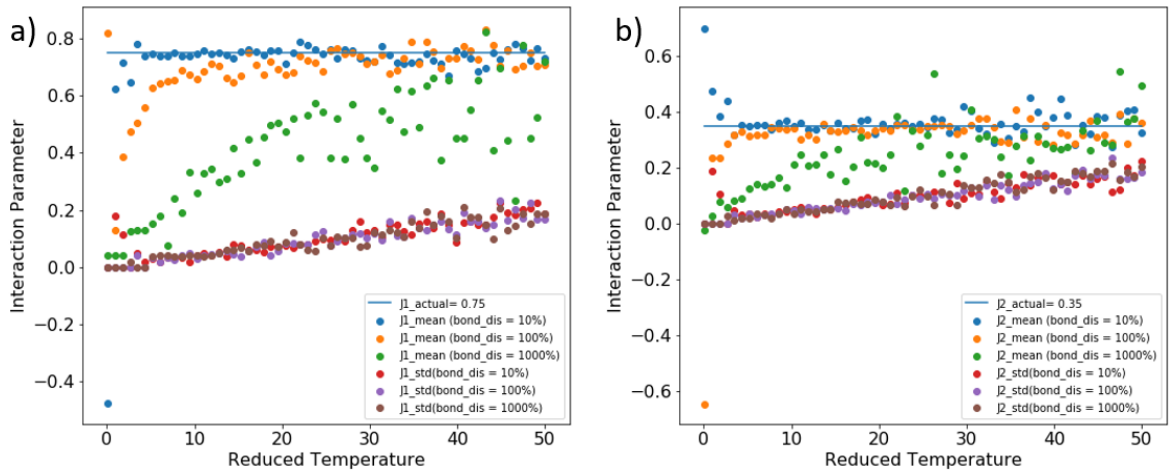

**Figure 8.** Reconstructed values of the exchange integrals and their uncertainties corresponding to the (a) nearest neighbor interactions, and (b) next nearest neighbor interactions as a function of reduced temperature and noise.

Reconstructions are performed for disorder levels of 10%, 100% and 1000% of the values of exchange integrals and the results are shown in Fig. 8. The reconstruction works well at 10%. With 100% disorder, the reconstruction starts to fail at the low temperatures as at these temperatures the values of exchange integrals have a very strong effect on configuration. The configurations get more random with increase in temperature and this is represented by increase in uncertainty in reconstructions. At 1000% disorder the reconstruction completely fails, and the proposed methods cannot be applied at these high levels of disorder.

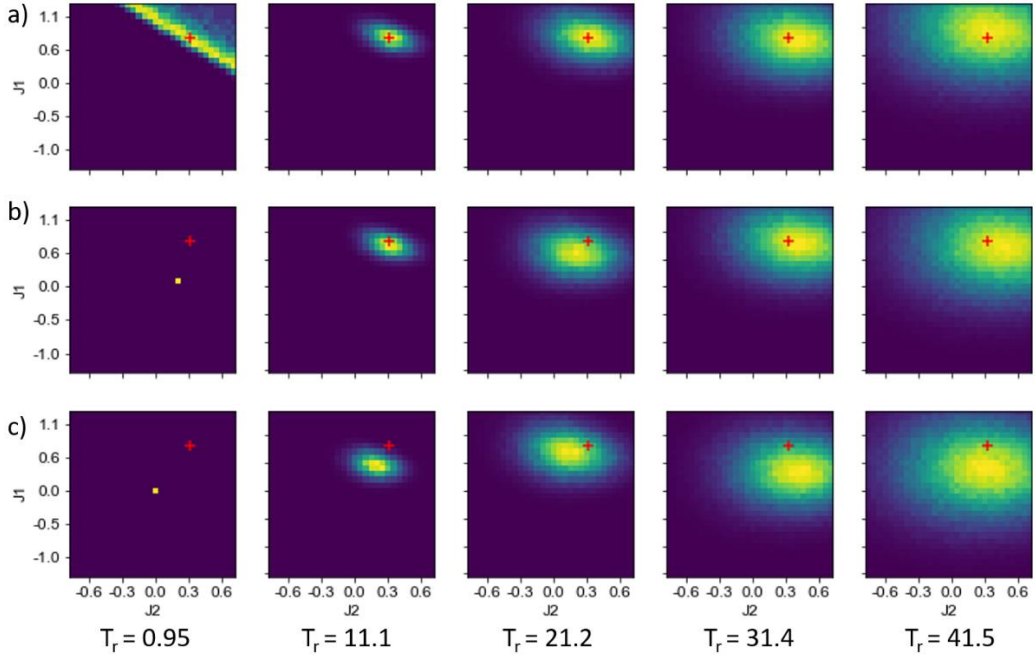

**Figure 9.** Likelihoods of the reconstructed values in the exchange integrals space at disorder levels of (a) 10%, (b) 100% and (c) 1000% at different points of reduced temperature. The values of the exchange integrals used to produce to the pseudo experimental data is marked by the cross in each plot.

The likelihood over the exchange integral space is shown in Fig. 9a-c at different points of reduced temperature. Note that for low temperatures below transition and low disorder the recovered exchange integrals are strongly correlated, suggesting that the system behavior belongs can be well defined by a single exchange integral. For higher disorder level reconstruction is impossible.

Above the transition, the reconstruction yields unbiased estimates of exchange integrals for low and intermediate disorder, whereas for high disorder the reconstructed integrals are below the true value. This trend continues for higher temperatures. Finally, in all cases the temperature increase yields broader distribution of posterior values, rendering reconstruction less reliable.

## 4. Conclusions

To summarize, here we have developed an approach for the reconstruction of the microscopic parameters of lattice model with two competing interactions from the observations of the microscopic degrees of freedom. We established the uncertainties and reliability of such analysis in a broad parameter-temperature space. Below the phase transition the reconstruction is impossible and can lead to the biased estimates of the parameters. However, these situations can be readily identified from experiment as the presence of large single-phase regions and correspondingly extremely poor sampling of the configuration space. At the same time, the reconstruction is possible well above the phase transition (1-2 order of magnitude) and in the regions of the parameter space when the macroscopic ground state of the system is poorly defined due to frustrated interactions. Similarly, the reconstruction is robust with respect to the frozen disorder. Interestingly, bond disorder tends to affect reconstruction close to transition temperatures leading to biased estimates of exchange integrals, whereas at high temperatures unbiased estimates with large uncertainty emerge.

As an ancillary task, we have developed a machine learning approach based on histogram clustering to predict phase diagrams efficiently using a reduced descriptor space. We illustrated that the phase evolution in the vicinity of regions comprising frustrated phase gives rise to the hierarchy of local configurations.

The explored approach can be applied to the traditionally complex problems of condensed matter physics such as ferroelectric relaxors and morphotropic phase boundary systems, spin and cluster glasses, quantum systems once the local descriptors linked to the relevant physical behaviors are known. Correspondingly, of interest becomes the role of precision in determination of local descriptors on parameter reconstruction, necessitating the development of appropriate Bayesian frameworks to link the experimental data and reconstructions.

**Acknowledgement**

This work is based upon work supported by the U.S. Department of Energy (DOE), Office of Science, Basic Energy Sciences (BES), Materials Sciences and Engineering Division (S.M.V., L.V., and S.V.K.) and was performed and partially supported (MZ, RKV) at the Oak Ridge National Laboratory's Center for Nanophase Materials Sciences (CNMS), a U.S. Department of Energy, Office of Science User Facility.